\documentclass[twocolumn,pre,superscriptaddress,notitlepage]{revtex4}

\usepackage{graphicx}
\usepackage{comment}
\usepackage{amsmath,amssymb,amsthm} 
\usepackage{verbatim}
\usepackage{float}

\setcitestyle{super}

\usepackage[sort&compress]{natbib}

\begin{document}

\title{The impact range for smooth wall-liquid interactions in nanoconfined liquids}
\author{Trond S. Ingebrigtsen}
\email{trond@ruc.dk}
\affiliation{DNRF Centre ``Glass and Time'', IMFUFA, Department of Sciences, Roskilde University, Postbox 260, DK-4000 Roskilde, Denmark}
\author{Jeppe C. Dyre}
\affiliation{DNRF Centre ``Glass and Time'', IMFUFA, Department of Sciences, Roskilde University, Postbox 260, DK-4000 Roskilde, Denmark}

\date{\today}

\begin{abstract}
  Bulk and nanoconfined liquids have initially very different physics; for instance, nanoconfined liquids show stratification and position-dependent 
  relaxation processes. A number of similarities between bulk and nanoconfined liquids have nevertheless been reported in computer simulations 
  during the last decade. Inspired by these observations, we present results from molecular dynamics computer simulations 
of three nanoconfined liquids (i.e., single-component Lennard-Jones (LJ) liquid, Kob-Andersen binary LJ mixture, and an asymmetric dumbbell model) 
  demonstrating also a microscopic similarity between bulk and nanoconfined liquids. The results show 
  that the interaction range for the wall-liquid and liquid-liquid interactions of the nanoconfined liquid are identical to the bulk liquid 
  as long as the liquid remains ''Roskilde simple'' in nanoconfinement, i.e., the liquid has strong correlations between virial and potential 
  energy equilibrium fluctuations in the \textit{NVT} ensemble. 
\end{abstract}

\maketitle

\section{Introduction}

Bulk and nanoconfined liquids have apparently very different physics. The confining boundaries not only bias the average
spatial distribution of the constituent molecules, but also the ways by which those molecules
can dynamically rearrange. These effects have been exploited in the 
design of coating, nanopatterning, and nanomanufacturing technologies\cite{bhushan1995,whitesides2006}. They have also 
been experimentally characterized for a wide variety of material systems, including 
small-molecule fluids\cite{drake1990,granick1991,morineau2002,jackson1991,teboul2005,alcoutlabi2005,coasne2011,richert2011}, 
polymers\cite{keddie1994,sergei2006,forrest2001,ellison2003,rittigstein2007,paeng2012}, 
ionic liquids\cite{iacob2012}, 
liquid crystals\cite{ji2009}, 
and dense colloidal suspensions\cite{nugent2007,eral2009,michailidou2009,watanabe2011,edmond2012}, as well as studied extensively 
via computer simulations\cite{fehr1995,torres2000,scheidler2000,starr2002,baljon2005,kurzidim2009,watanabe2011,starr2011,betancourt2013}.  

A number of similarities between bulk and nanoconfined liquids have nevertheless been observed during the last decade. For example, it has been 
discovered that Rosenfeld's excess entropy scaling\cite{rosenfeld1} $-$ in which a dimensionless 
transport coefficient is correlated with the excess entropy (with respect to an ideal gas) $-$ is unaffected  
by the spatial confinement\cite{hsconfined,dumbbellconfinedroughwalls,goel2009,ingebrigtsen2013}. This also turns out to be 
valid for excess isochoric heat capacity scaling\cite{ingebrigtsen2013} and for Rosenfeld-Tarazona's expression for the 
specific heat\cite{RT,ingebrigtsen2013b}. Similarly, it has been shown that the continuum Navier-Stokes equations extend into the nanoscale regime 
by the inclusion of
coupling between linear and angular momentum\cite{alley1983,travis1997,hansen2007,hansen2013}. Likewise have attempts at extending bulk 
mode-coupling theory\cite{gotze2008} to the nanoscale been fruitful\cite{krakoviack2005,biroli2006,lang2012}. These results are
naturally very intriguing since they indicate that fundamentally new theories may not be needed to describe the 
physics of the nanoscale. At the same time, the results are also quite puzzling given the fact that bulk and 
nanoconfined liquids have very different phenomenology.

In this connection, it is thus natural to wonder whether these similarities 
extend to a more fundamental or ''microscopic'' level. In fact, the mere difference between bulk and nanoconfined liquids is the external potential constituting the 
spatial confinement. The study of 
confinement is nevertheless complicated by the fact that the confining boundaries (the ''walls'') do not only impose a geometrical 
constraint on the liquid, but also energetic costs. Several investigations have tried to separate the entropic and energetic 
contributions to the physics by studying, e.g., hard-sphere walls\cite{diestler1991}, reflecting walls\cite{krekelberg2011}, 
amorphous walls\cite{scheidler2000,scheidler2000b}, and more. With the above mentioned similarities in mind,  a ''microscopic'' conjecture motivated from these studies is 
that the interaction range for the wall-liquid and liquid-liquid interactions of the nanoconfined liquid is identical to that of the bulk liquid.

The relevant range of interaction for bulk liquids was considered long ago by traditional liquid-state theories\cite{widom1967,wca} and is 
captured in the well-known Weeks-Chandler-Andersen (WCA) approximation\cite{wca} in which the pair potential is truncated at the 
potential minimum. In this picture, the attractive pair forces of typical liquid-state configurations cancel to a good approximation 
and only contribute with a negative background potential\cite{widom1967}. The relevance of a sharp 
distinction between repulsive and attractive pair forces of the bulk liquid 
has, however, recently been questioned\cite{FCS2,prx,bohling2013}. Instead, it has been suggested that the relevant 
distinction is between the forces within the first coordination shell (FCS) of molecules and those outside the FCS\cite{prx}.

The traditional bulk liquid-state picture has, in fact, for nonuniform liquids been questioned\cite{weeks1995}; in particular, 
near the confining walls and 
close to drying transitions. Several promising theories\cite{weeks1995,weeks1997,weeks1998,rodgers2008,wang2012}, such as Local Molecular Field (LMF)
theory\cite{weeks1995,weeks1997,weeks1998,rodgers2008}, have emerged to take into account the importance of the attractive forces in nonuniform liquids.
We study here whether the FCS picture (i.e., that interactions beyond the FCS can safely be ignored) of the bulk liquid state holds when considering nanoconfined liquids, in 
particular, for the wall-liquid interactions. If this turns out to be the case for all liquids or perhaps just for a well-defined class of liquids, 
a more fundamental or microscopic similarity between bulk and nanoconfined liquids is hereby established. 

\section{Methods}

We apply in the current study GPU-optimized \textit{NVT} molecular dynamics computer 
simulations\cite{nvttoxvaerd,toxconstraintnve} (http://rumd.org) of several 
atomic and molecular model systems in nanoconfinement. 
The confinement is modelled as a symmetric slit-pore geometry using a smooth external potential\cite{ingebrigtsen2013}. 
More specifically, the external wall potential is given by

\begin{equation}
  v_{9,3}(z) = \frac{4\pi\epsilon_{iw}\rho_{w}\sigma^{3}_{iw}}{3}\Big[\frac{1}{15}\Big(\frac{\sigma_{iw}}{z}\Big)^{9} - \frac{1}{2}\Big(\frac{\sigma_{iw}}{z}\Big)^{3}\Big]. \label{steele}
\end{equation}
\newline
Here, $z$ is the distance between the divergence of the potential and the particle in question. $\sigma_{iw}$ and $\epsilon_{iw}$ are parameters similar to the 
Lennard-Jones (LJ) potential, and $\rho_{w}$ defines the density of the confining solid. Equation (\ref{steele}) 
appears after considering the total interaction of a LJ particle with a 
semi-infinite solid continuum of LJ particles \cite{steele1973} and has been used in some of the earliest studies of 
confinement\cite{tox1981,schoen1987}. Quantities are here and henceforth reported in dimensionless LJ units by setting $\sigma_{AA} = 1$, 
$\epsilon_{AA} = 1$, etc.
We use wall-parameters $\sigma_{Aw}$ = 1, $\epsilon_{Aw}$ = 1; in the case of binary systems $\sigma_{Bw}$ = $(1 + \sigma_{BB})/2$, $\epsilon_{Bw}$ = $\sqrt{ 1 \cdot \epsilon_{BB}}$. The density of the confining solid $\sigma_{w}$ is in most cases
chosen to be equal to the total average slit-pore density\cite{ingebrigtsen2013}.

However, to perform simulations with FCS cutoffs it has recently been shown\cite{FCS1} that ensuring the continuity of forces is of utmost importance. 
One method of achieving a continuous force is the so-called shifted-force (SF) cutoff 

\begin{equation}
  f_{\rm SF}(r)\,=\,
  \begin{cases}  f(r)-f(r_c) & \text{if}\,\, r<r_c\,, \\ 
    0 &\text{if}\,\, r>r_c\,,
  \end{cases} 
\end{equation}
which corresponds to adding a 
linear term $-v(r_{c}) ( r - r_{c} ) - v(r_{c})$ to the original potential $v(r)$. In the bulk liquid, these additional terms 
sum to a good approximation to a constant, and thus do not affect the structure nor the dynamics\cite{paper2,prx}. For 
the linear term of the liquid-liquid interactions to sum to a constant also near the walls of the confinement, we have 
found that an additional term beyond the SF-correction on the wall-liquid interactions is needed. We have observed that 
adding 

\begin{equation}
\Delta f = |f_{PAIR}(r_{ll,FCS}) - f_{WALL}(r_{wl,FCS})|,
\end{equation}
to the force from the wall (restricted to the FCS) often gives a good cancellation of the linear 
term near the walls. Here $r_{ll,FCS}$ and  $r_{wl,FCS}$ are, respectively, the distance of the FCS for the liquid-liquid (ll) and 
wall-liquid (wl) interactions (see next section for the FCS delimitation) and are thus not free parameters. In the case of 
different particle-particle interactions we apply cutoffs in units of the largest particle. Applying this particular cutoff method 
to all FCS simulations, we now present results using FCS cutoffs in nanoconfined liquids.

\section{Results}

\begin{figure}[H]
  \centering
  \includegraphics[width=80mm]{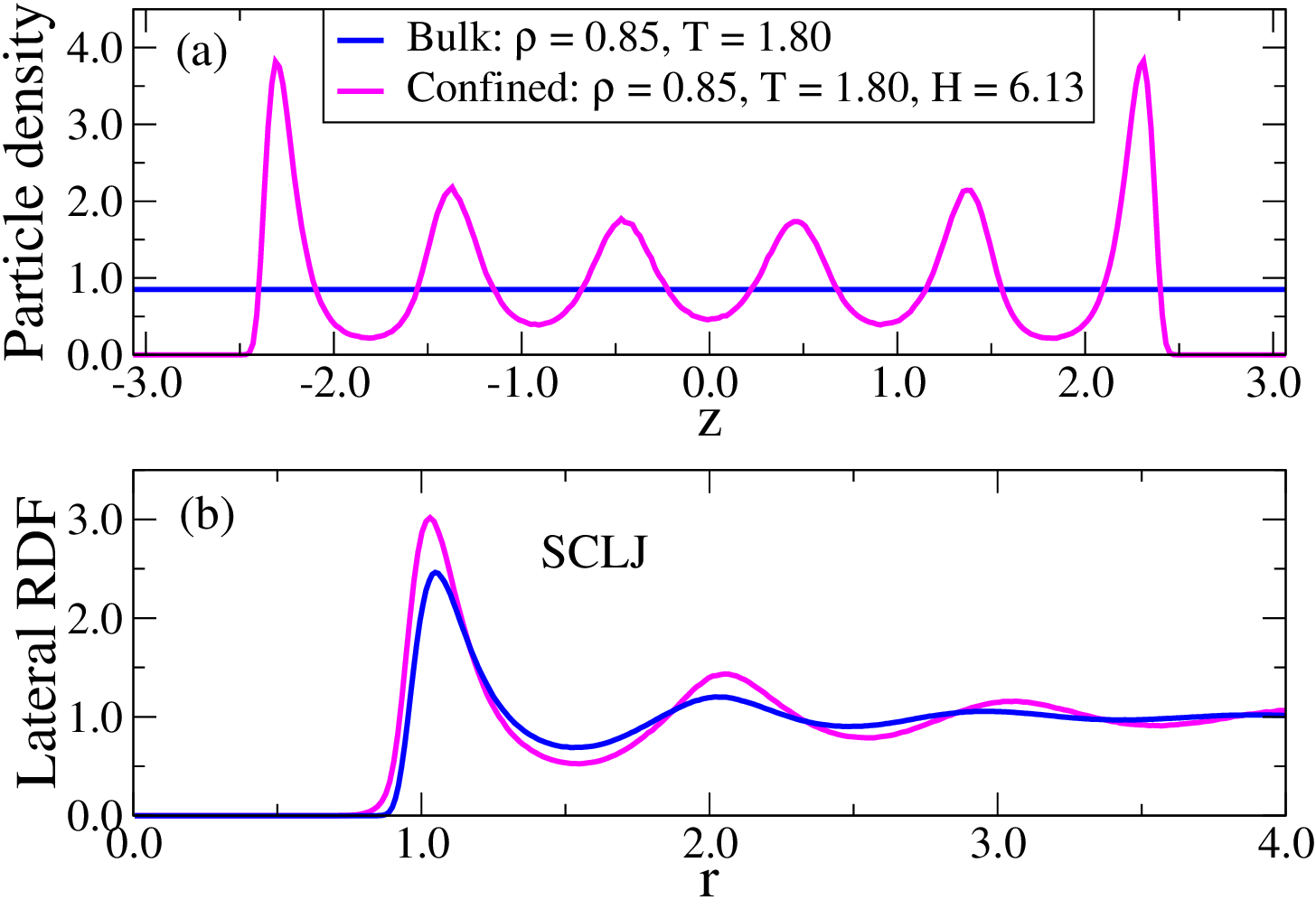}
  \caption{Impact of the slit-pore confinement (magenta) on the bulk single-component 
    Lennard-Jones (SCLJ) liquid (blue). (a) Density profiles. (b) Lateral RDFs. The RDF 
    in the nanoconfinement is calculated for the contact layer (i.e., the layer closest to the walls) showing a  
    distinct structure from the bulk liquid.}
  \label{FCSCON1}
\end{figure}

The investigation is started with Fig. \ref{FCSCON1} showing the impact of the slit-pore confinement (magenta) on the bulk single-component
Lennard-Jones (SCLJ) liquid (blue). 
Figure \ref{FCSCON1}(a) shows the density profile whereas Fig. \ref{FCSCON1}(b) shows the lateral radial distribution function (RDF). The RDF in the nanoconfinement is calculated for the 
contact layer (i.e., the layer closest to the walls), and the comparison is performed at the same average density and temperature. The slit-pore has a significant impact on the structure of the liquid showing large density oscillations near the walls, which is also manifested in the RDF obtained near the walls. The contact-layer RDF is quite distinct from the bulk liquid by showing a higher degree of ordering 
in the liquid. The latter confirms results from other computer simulations of nanoconfined liquids\cite{schoen1987,watanabe2011}.\newline \newline

From these observations it appears that the microscopic physics obtained near the walls is actually rather different from 
the bulk liquid. The latter is also to be expected due to, in general, diverse and complicated wall-liquid interactions. 

\subsection{Simulations of FCS cutoffs}

\begin{figure}[H]
  \centering
  \includegraphics[width=80mm]{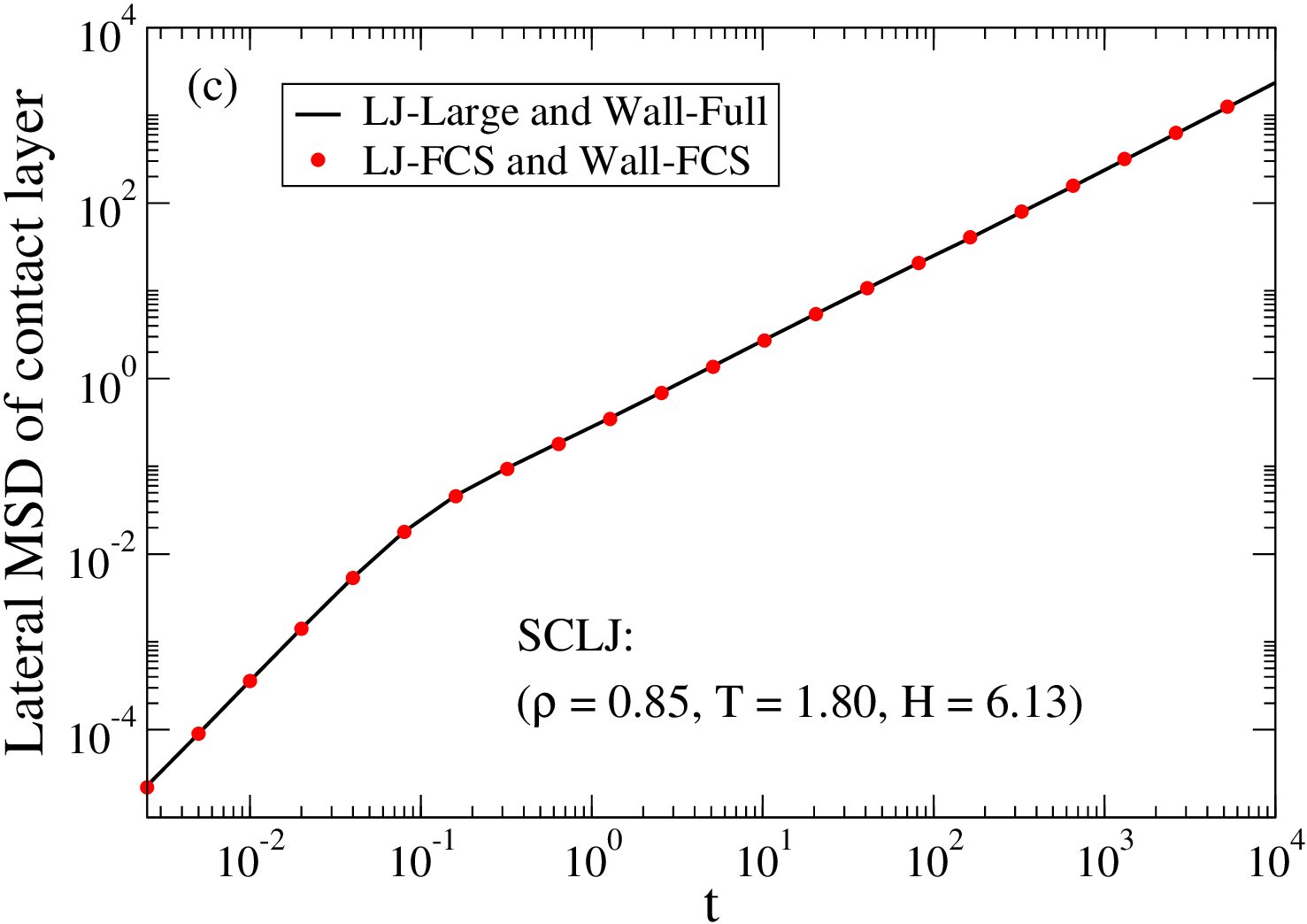}
  \caption{FCS simulations for the SCLJ liquid in a slit-pore at $\rho$ = 0.85, $T$ = 1.80, $H$ = 6.13. The black curves 
    give a simulation with a large pair potential cutoff in a combination with no cutoff for the wall-liquid interactions. The red dots 
    give results for a cutoff at the FCS for the liquid-liquid and wall-liquid interactions ($r_{ll,FCS} = 1.549$ and $r_{wl,FCS}$ = 1.250). $R$ is 
    the correlation coefficient\cite{paper1} 
    between \textit{NVT} virial and potential energy equilibrium fluctuations and is defined in Eq. (\ref{rcor}) below. (a) Density profiles. 
    (b) Lateral RDFs of the contact layer. (c) Lateral mean-square displacements (MSD) of the contact layer.}
  \label{FCSCON2}
\end{figure}

We now proceed to test the bulk FCS picture in nanoconfinement. Figure \ref{FCSCON2} shows density profiles and contact-layer RDFs for the 
SCLJ confined state point of Fig. \ref{FCSCON1}. The black curves give a simulation with a 
large pair potential cutoff in a combination with no cutoff for the wall-liquid interactions. The red dots give results for an FCS cutoff 
for the liquid-liquid and wall-liquid interactions. The FCSs are delimited\cite{prx} via the first minimum of the contact-layer RDF and density profile, respectively.
The FCS cutoffs  are seen to capture the relevant physics for the entire slit-pore even though the walls modify the structure of liquid quite significantly with 
respect to the bulk liquid (Fig. \ref{FCSCON1}(b)).\newline \newline

\begin{figure}[H]
  \centering
  \includegraphics[width=80mm]{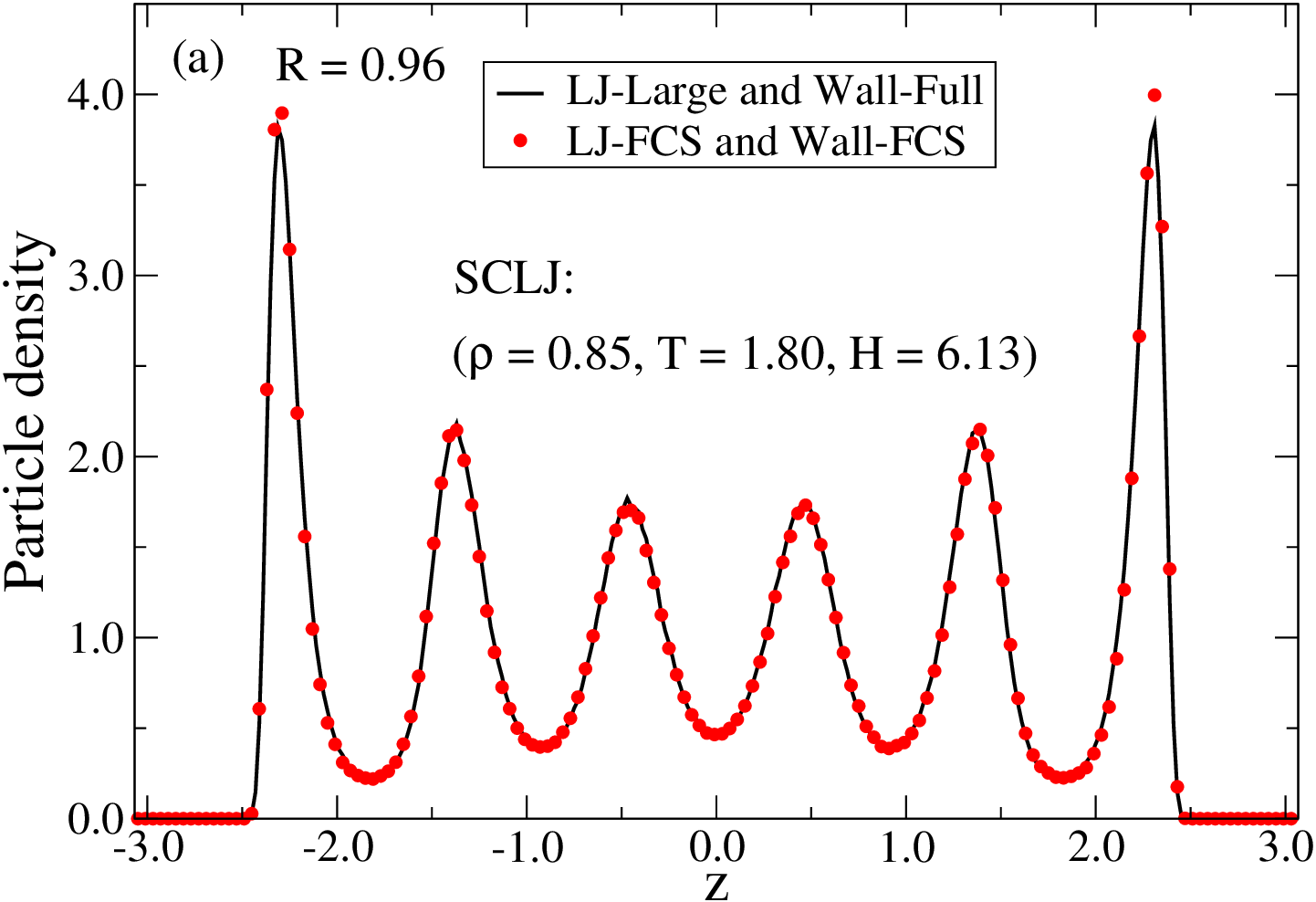}
\end{figure}
\begin{figure}[H]
  \centering
  \includegraphics[width=80mm]{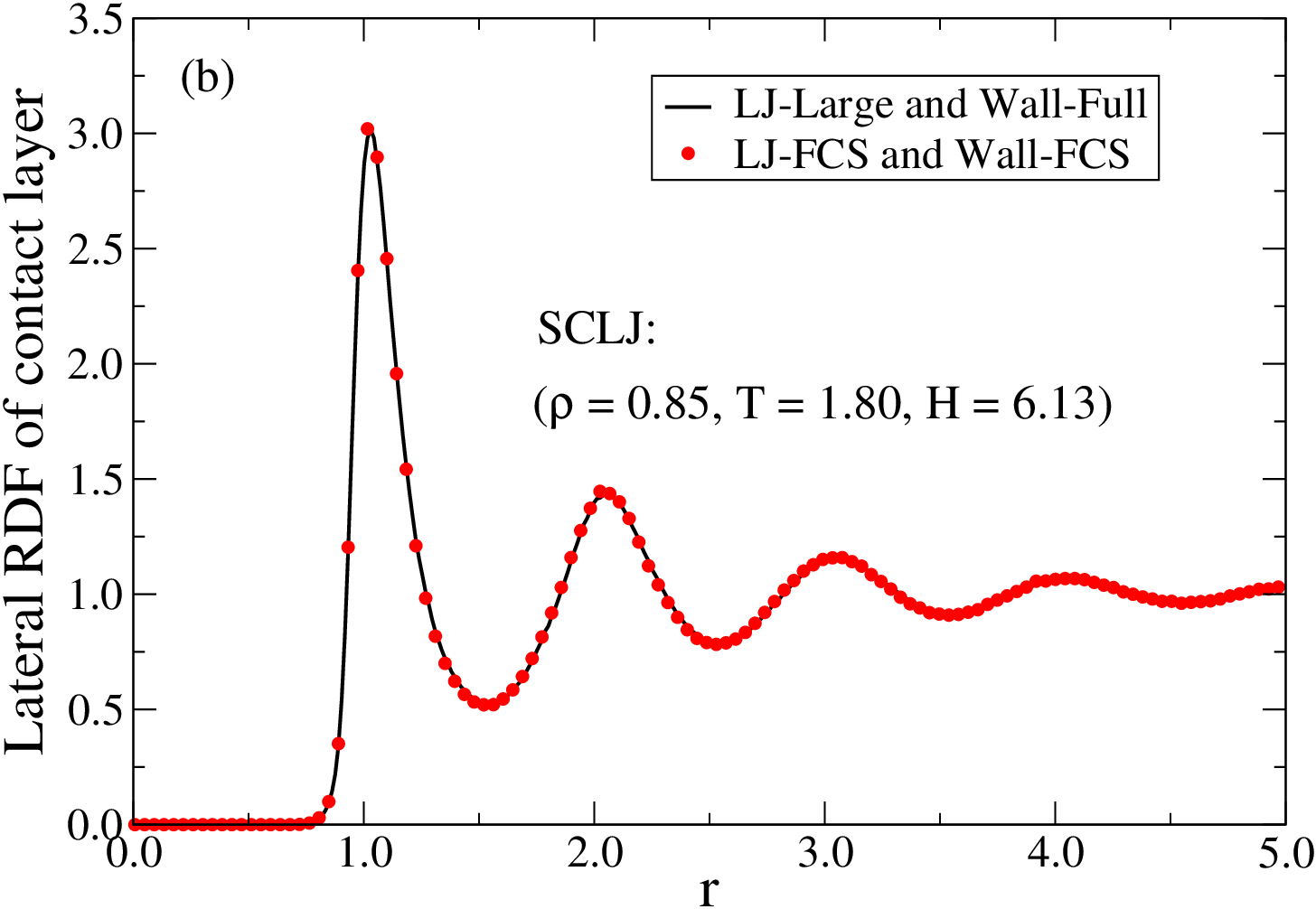}
\end{figure}
\begin{figure}[H]
  \centering
  \includegraphics[width=80mm]{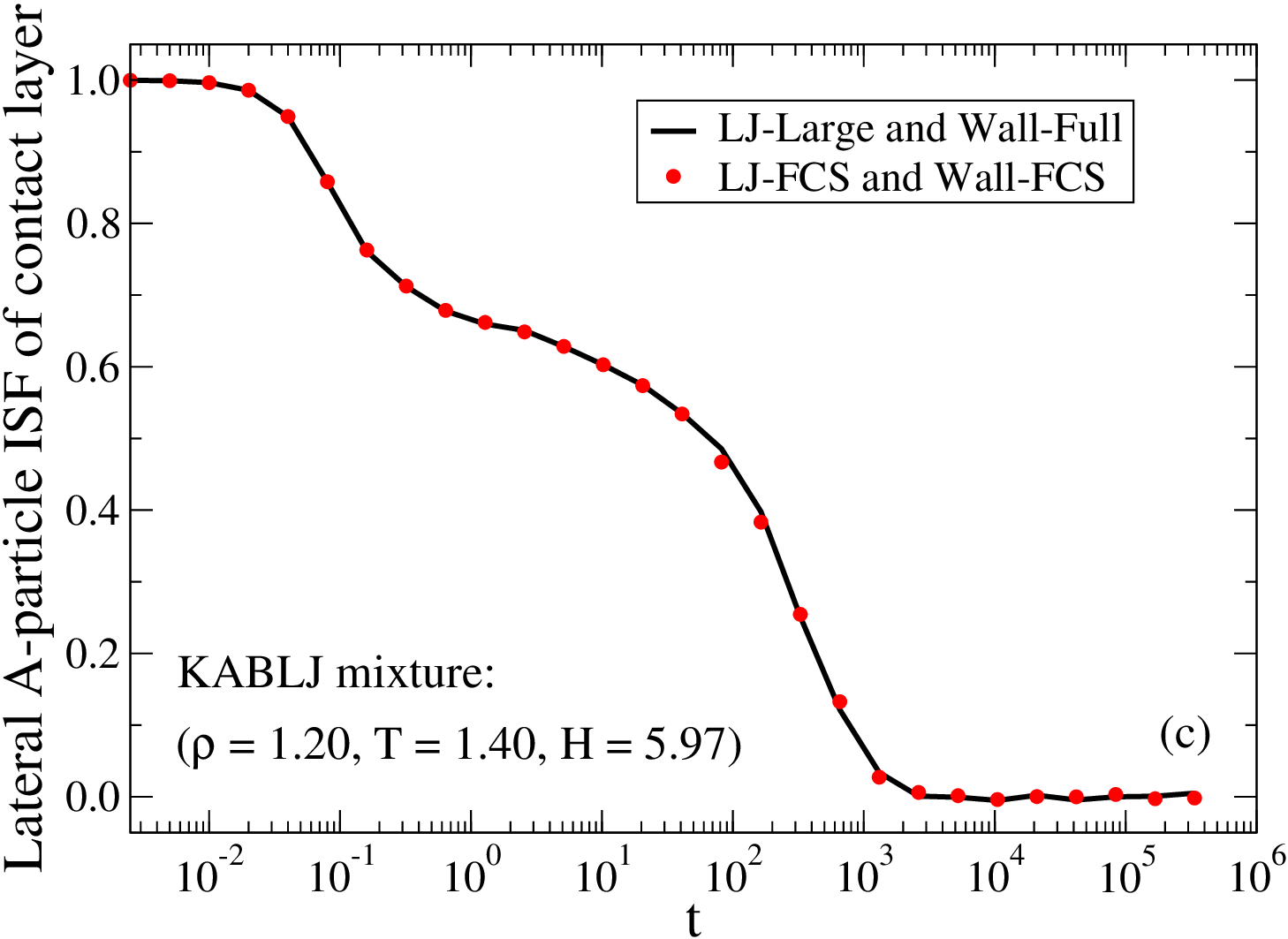}
  \caption{FCS simulations for the KABLJ mixture  in a slit-pore at $\rho$ = 1.20, $T$ = 1.40, $H$ = 5.97 ($r_{ll,FCS}$ = 1.479 and $r_{wl,FCS}$ = 1.252). 
    $R$ is defined in Eq. (\ref{rcor}).
    (a) Density profiles. (b) Lateral RDFs of the contact layer. The BB-particle RDF has more noise than the other particle-particle RDFs. (c) Lateral $A$-particle ISFs of the contact layer.}
  \label{FCSCON3}
\end{figure}

To test the FCS picture further for nanoconfined liquids, Fig. \ref{FCSCON3} shows equivalent simulations for the Kob-Andersen binary LJ mixture\cite{ka1} (KABLJ) at $\rho$ = 1.20, $T$ = 1.40, $H$ = 5.97. 
Figure \ref{FCSCON3}(a) shows $A$-particle density profiles, Fig. \ref{FCSCON3}(b) shows lateral RDFs of the contact layer, and 
Fig. \ref{FCSCON3}(c) shows lateral $A$-particle incoherent intermediate scattering functions (ISFs) of the 
contact layer. The physics near the walls is seen to be captured excellently using the FCS cutoffs.\newline \newline

\begin{figure}[H]
  \centering
  \includegraphics[width=80mm]{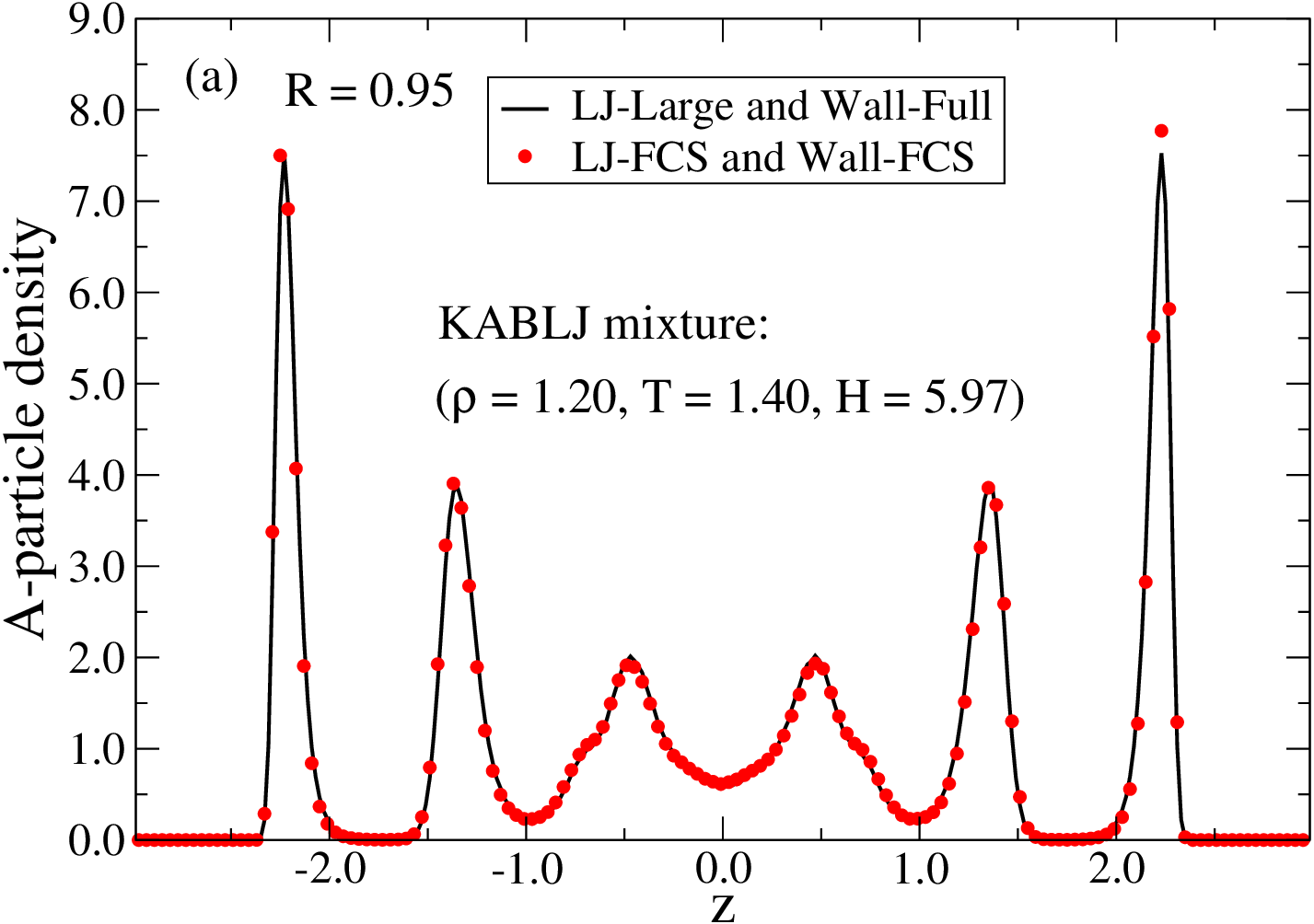}
\end{figure}
\begin{figure}[H]
  \centering
  \includegraphics[width=80mm]{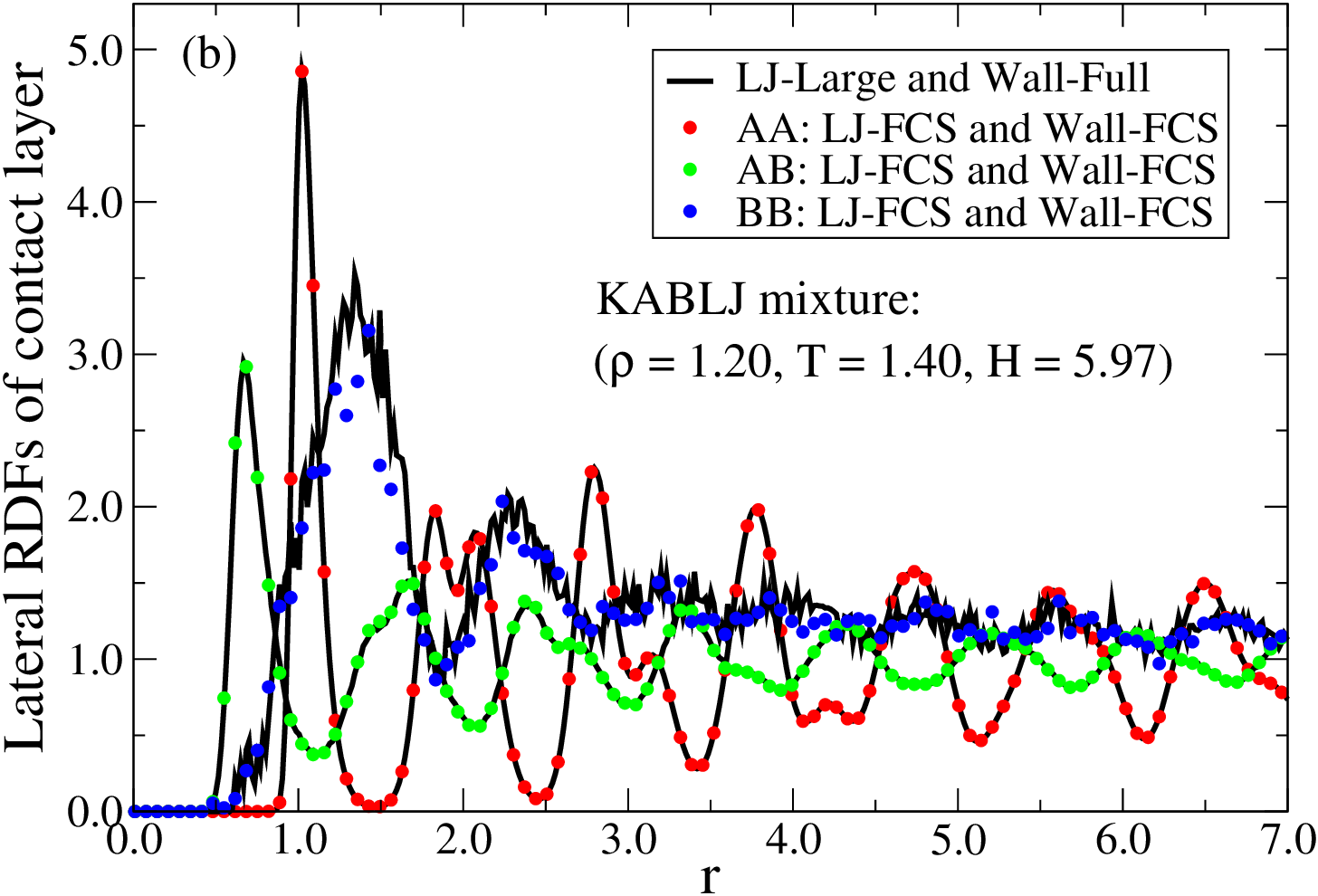}
\end{figure}
\begin{figure}[H]
  \centering
  \includegraphics[width=80mm]{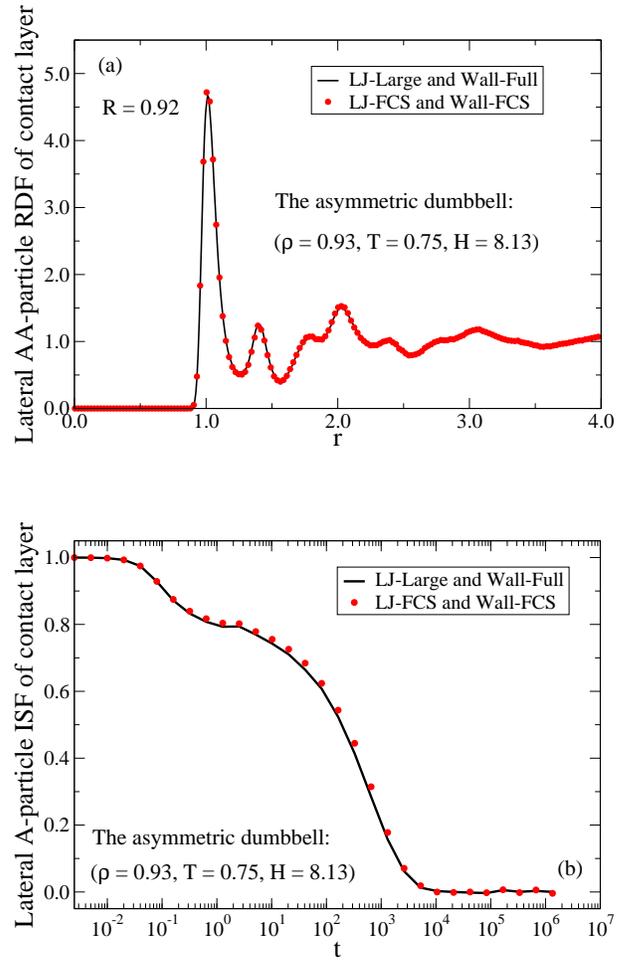}
  \caption{FCS simulations for the asymmetric dumbbell model in a slit-pore at $\rho$ = 0.93, $T$ = 0.75, $H$ = 8.13 
    ($r_{ll,FCS}$ = 1.565 and $r_{wl,FCS}$ = 1.015). $R$ is defined in Eq. (\ref{rcor}). (a) Lateral RDFs of 
    the contact layer. (b) Lateral ISFs of the contact layer.}
  \label{FCSCON7}
\end{figure}

Proceeding to study in Fig. \ref{FCSCON7}, a small-molecule liquid in terms of the asymmetric dumbbell model\cite{moleculeshidden}, we 
show lateral contact-layer RDFs and ISFs. Slight deviations are seen with the FCS cutoff for 
the ISF and are assumed to be attributed to the approximative linear-cancellation term added to the force from the walls.\newline \newline

\begin{figure}[H]
  \centering
  \includegraphics[width=80mm]{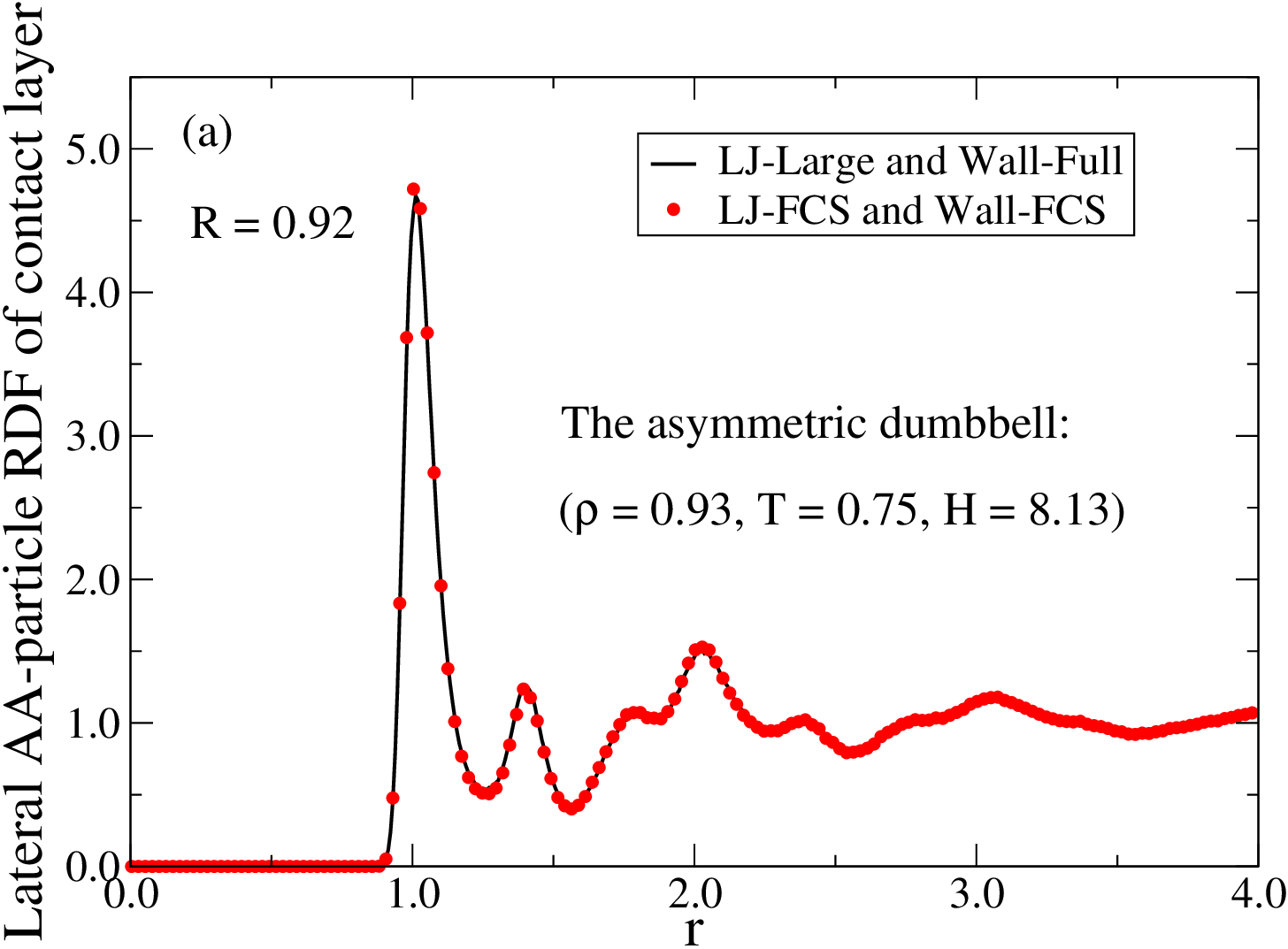}
\end{figure}
One may wonder how general are these results? The FCS interactions of the bulk liquid were shown in Ref. \onlinecite{prx} as the 
relevant interactions only when the 
liquid is so-called ''Roskilde simple''. Roskilde liquids \cite{paper1} are defined by having the \textit{NVT} virial/potential 
energy correlation coefficient (which depends on the state point),

\begin{equation}
R = \frac{\langle \Delta W \Delta U \rangle}{\sqrt{\langle (\Delta W)^{2} \rangle}\sqrt{\langle (\Delta U)^{2} \rangle}},\label{rcor}
\end{equation}
greater than $R \geq 0.90$. Only inverse power-law fluids are perfectly correlating ($R$ = 1), but many models\cite{paper1,prx} (e.g., 
the SCLJ liquid, KABLJ mixture, Lewis-Wahstr{\"o}m OTP\cite{otp1}, and more) as well as some 
experimental liquids\cite{gammagamma,roed2013} have been shown to belong to the class of Roskilde liquids. Roskilde liquids include most 
or all van der Waals and metallic liquids, whereas covalently, hydrogen-bonding or strongly ionic or dipolar liquids are generally 
not Roskilde simple\cite{paper1}. The latter reflects the fact that directional interactions tend to destroy the strong virial/potential energy correlation.  

\begin{figure}[H]
  \centering
  \includegraphics[width=80mm]{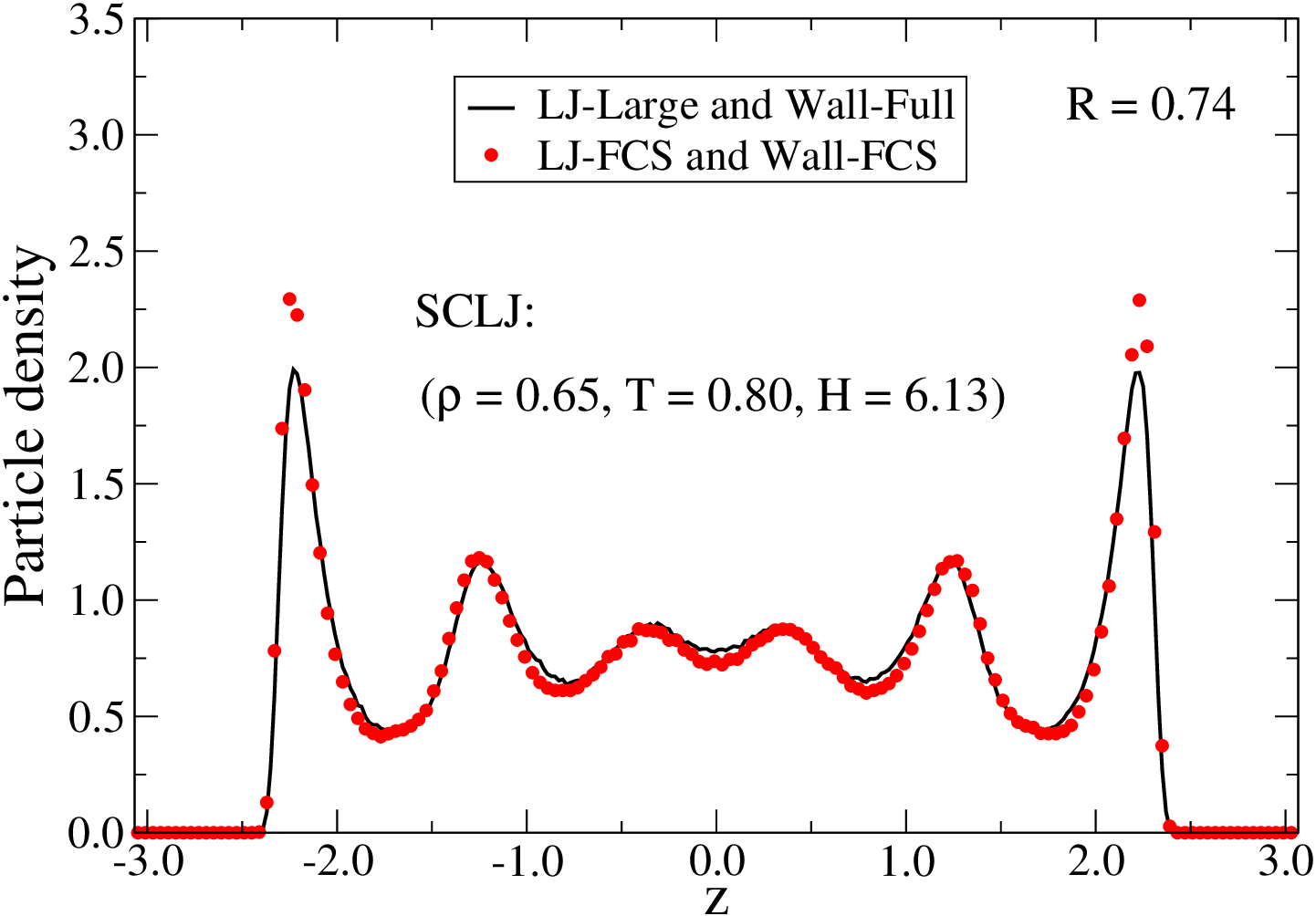}
  \caption{FCS simulations for the SCLJ liquid in a slit-pore at $\rho$ = 0.65, $T$ = 0.80, $H$ = 6.13 
    ($r_{ll,FCS}$ = 1.660 and $r_{wl,FCS}$ = 1.377). $R$ is defined in Eq. (\ref{rcor}). At this state point, the SCLJ liquid is not Roskilde simple with $R = 0.74$.}
  \label{FCSCON5}
\end{figure}

In all examples studied so far the liquid exhibits strong virial/potential energy correlation in confinement, too ($R \geq$ 0.90; see figures). 
It is possible that as the virial/potential energy 
correlation decreases, the FCS picture of the nanoconfined liquid becomes a worse approximation. To test this conjecture, we now turn the study to non-Roskilde liquids. One possible way of 
obtaining a non-Roskilde liquid is 
to decrease the density for the SCLJ liquid (recall that $R$ depends on state point). Doing so, we obtain the results of Fig. \ref{FCSCON5}; the correlation coefficient is here $R = 0.74$. We observe  
larger deviations with the FCS cutoff than previously noted and is consistent with the above conjecture.\newline \newline

\begin{figure}[H]
  \centering
  \includegraphics[width=80mm]{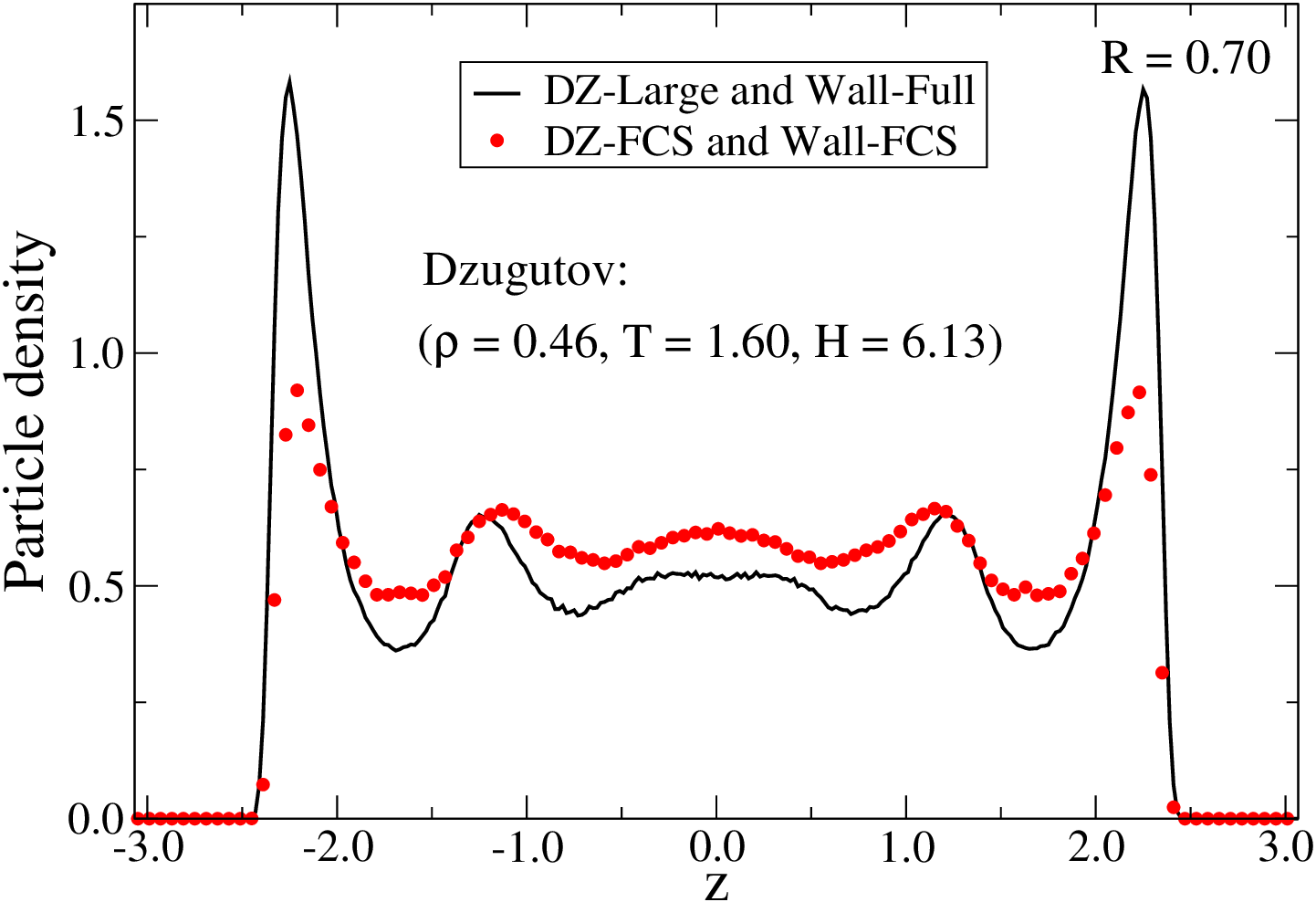}
  \caption{FCS simulations for the Dzugutov liquid in a slit-pore at $\rho$ = 0.46, $T$ = 1.60, $H$ = 6.13 
    ($r_{ll,FCS}$ = 1.616 and $r_{wl,FCS}$ = 1.406). $R$ is defined in Eq. (\ref{rcor}). 
    The Dzugutov liquid is not a Roskilde liquid with $R = 0.70$ at this state point.}
  \label{FCSCON4}
\end{figure}

As an additional example, we show a simulation in Fig. \ref{FCSCON4} of the Dzugutov liquid\cite{dzugutov} which is also a 
non-Roskilde liquid with $R = 0.70$. Here, the failure of the FCS cutoff in nanoconfinement is rather drastic.\newline \newline

\subsection{Simulations of WCA cutoffs}

\begin{figure}[H]
  \centering
  \includegraphics[width=80mm]{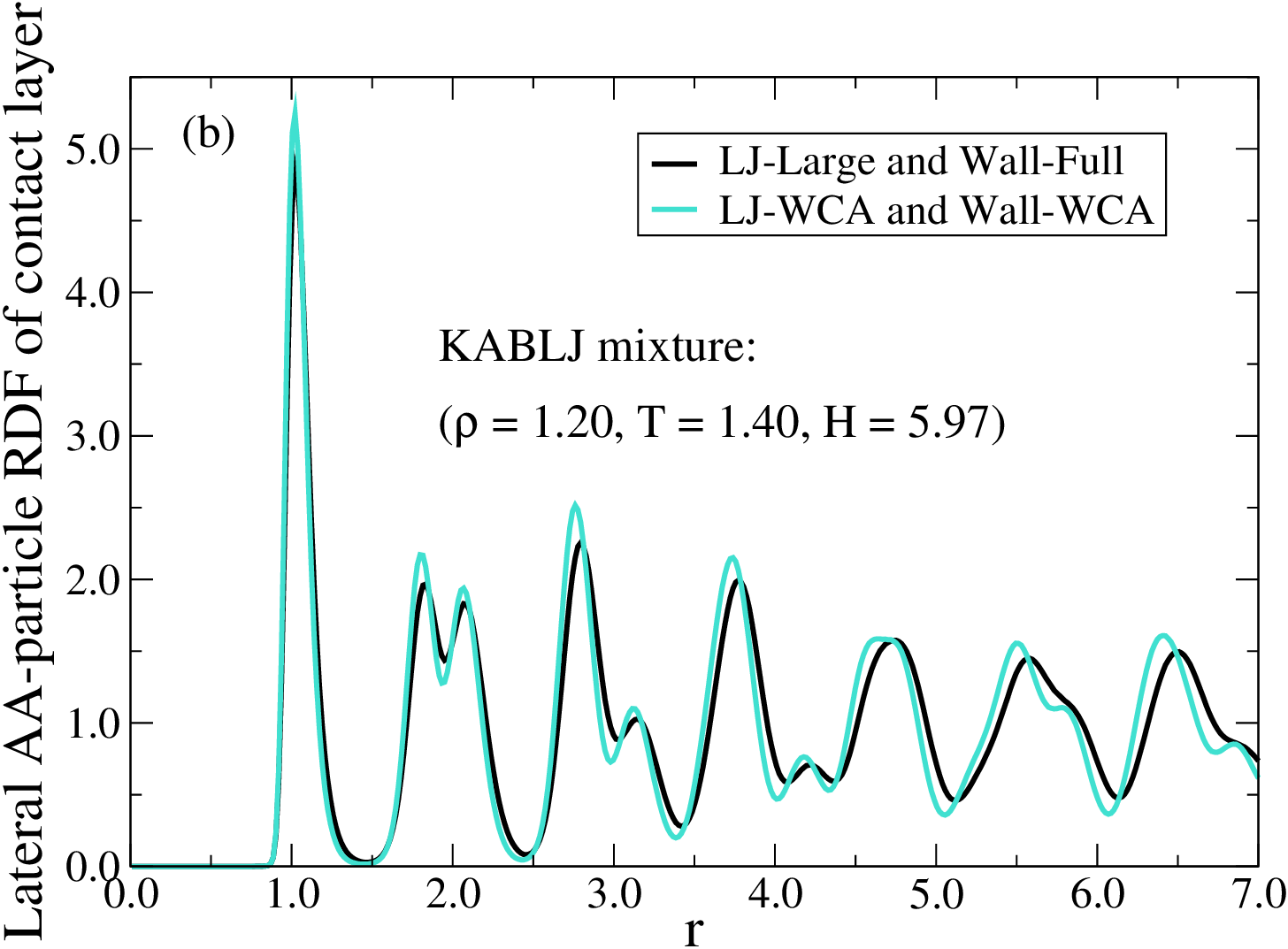}
  \caption{Simulations of the KABLJ mixture confined to a slit-pore at $\rho$ = 1.20, $T$ = 1.40, $H$ = 5.97. The WCA-cutoff 
    method is applied to both the liquid-liquid and wall-liquid interactions. (a) $A$-particle density profiles. (b) Lateral $AA$-particle 
    RDFs of contact layer.}
  \label{FCSCON8}
\end{figure}

Having established that for Roskilde liquids it is possible to safely ignore the interactions beyond the FCS for both the liquid-liquid and wall-liquid 
interactions, we now consider the effect of decreasing the cutoff below the FCS. 
According to the WCA philosophy in the bulk liquid\cite{wca}, it should be 
possible to cut the potentials at the potential minima and still obtain the correct physics. Applying this method
to simulations of confined liquids, we obtain the results of Fig. \ref{FCSCON8} for the KABLJ mixture. We observe 
significant discrepancy with the reference simulation hereby confirming results of previous simulations of confined liquids\cite{weeks1995,weeks1997}.\newline \newline

\begin{figure}[H]
  \centering
  \includegraphics[width=80mm]{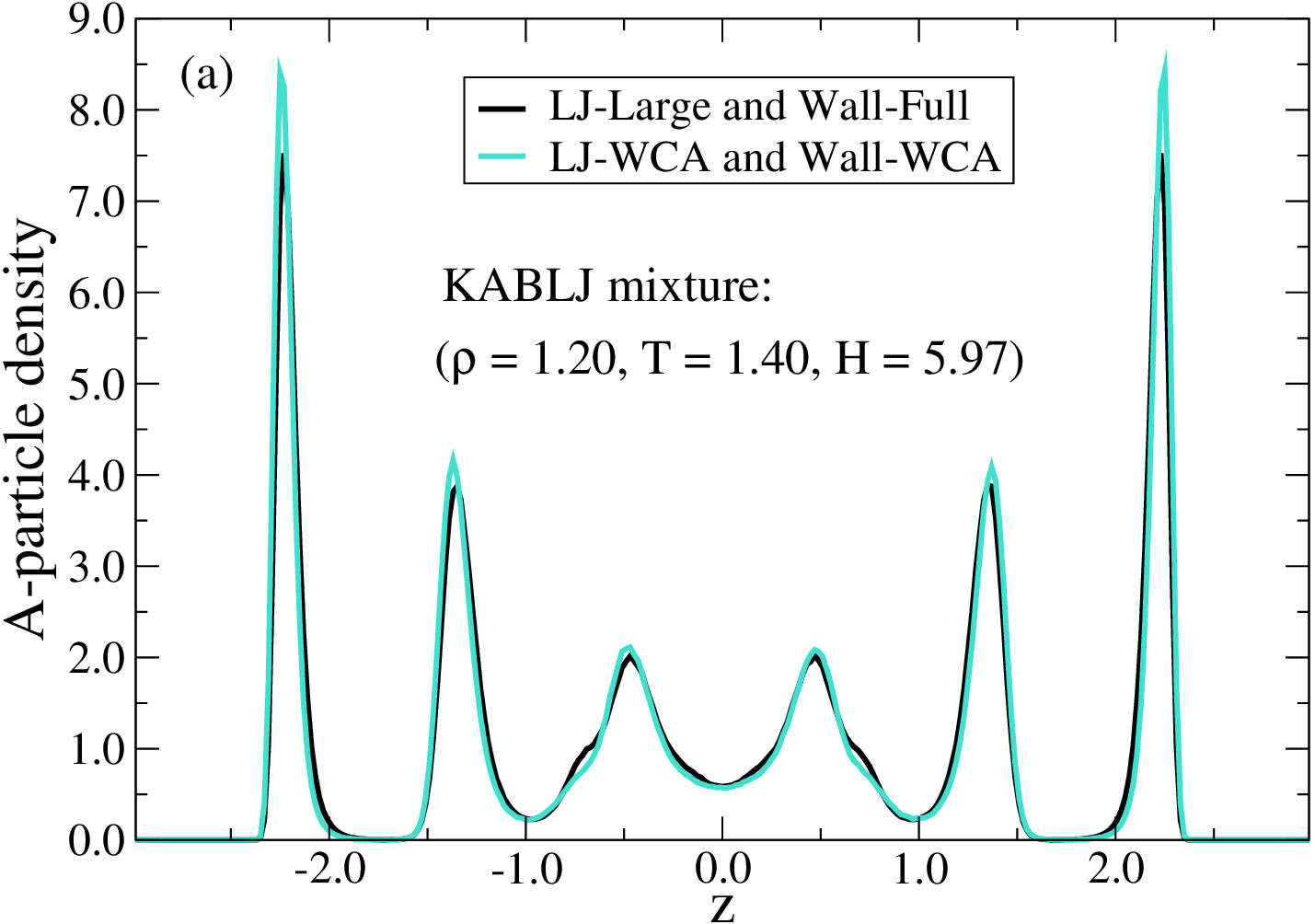}
\end{figure}
\begin{figure}[H]
  \centering
  \includegraphics[width=80mm]{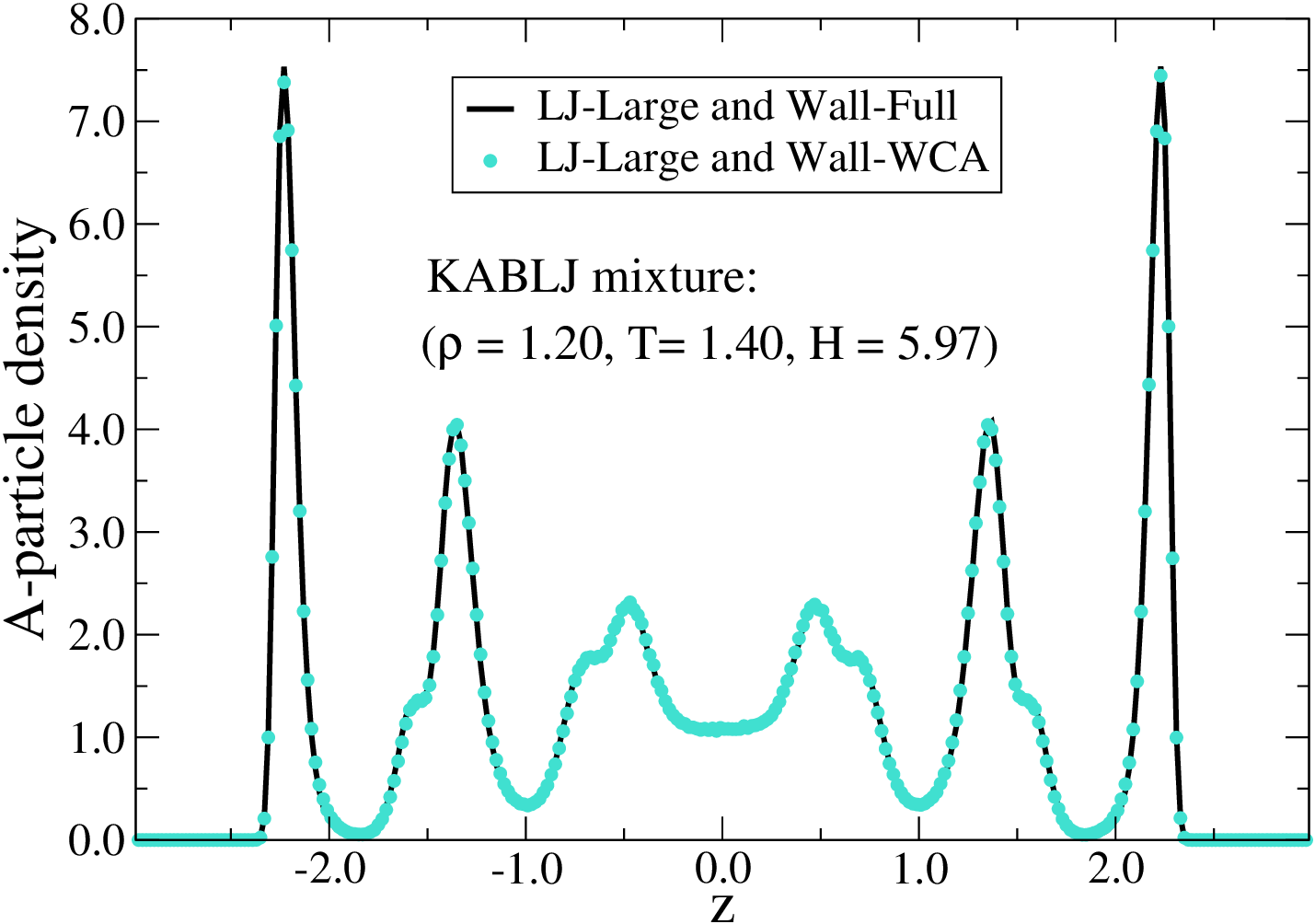}
  \caption{Simulations of the KABLJ mixture confined to a slit-pore at $\rho$ = 1.20, $T$ = 1.40, $H$ = 5.97. The WCA-cutoff 
    method is applied only to the wall-liquid interactions.}
  \label{FCSCON9}
\end{figure}

The discrepancy is noted to be the largest near the walls, and to understand this behavior in more detail, we show results in 
Fig. \ref{FCSCON9} applying only the WCA approximation on the wall-liquid interactions. Here, no discrepancy is noted.\newline \newline

From these results one might be tempted to conclude that the FCS is not the fundamental distance for the wall-liquid interactions. The distance of 
the FCS does, however, depend on density and thus it is possible to perform simulations in which the location of the FCS is pushed far to the left 
of the potential minimum. 

\begin{figure}[H]
  \centering
  \includegraphics[width=80mm]{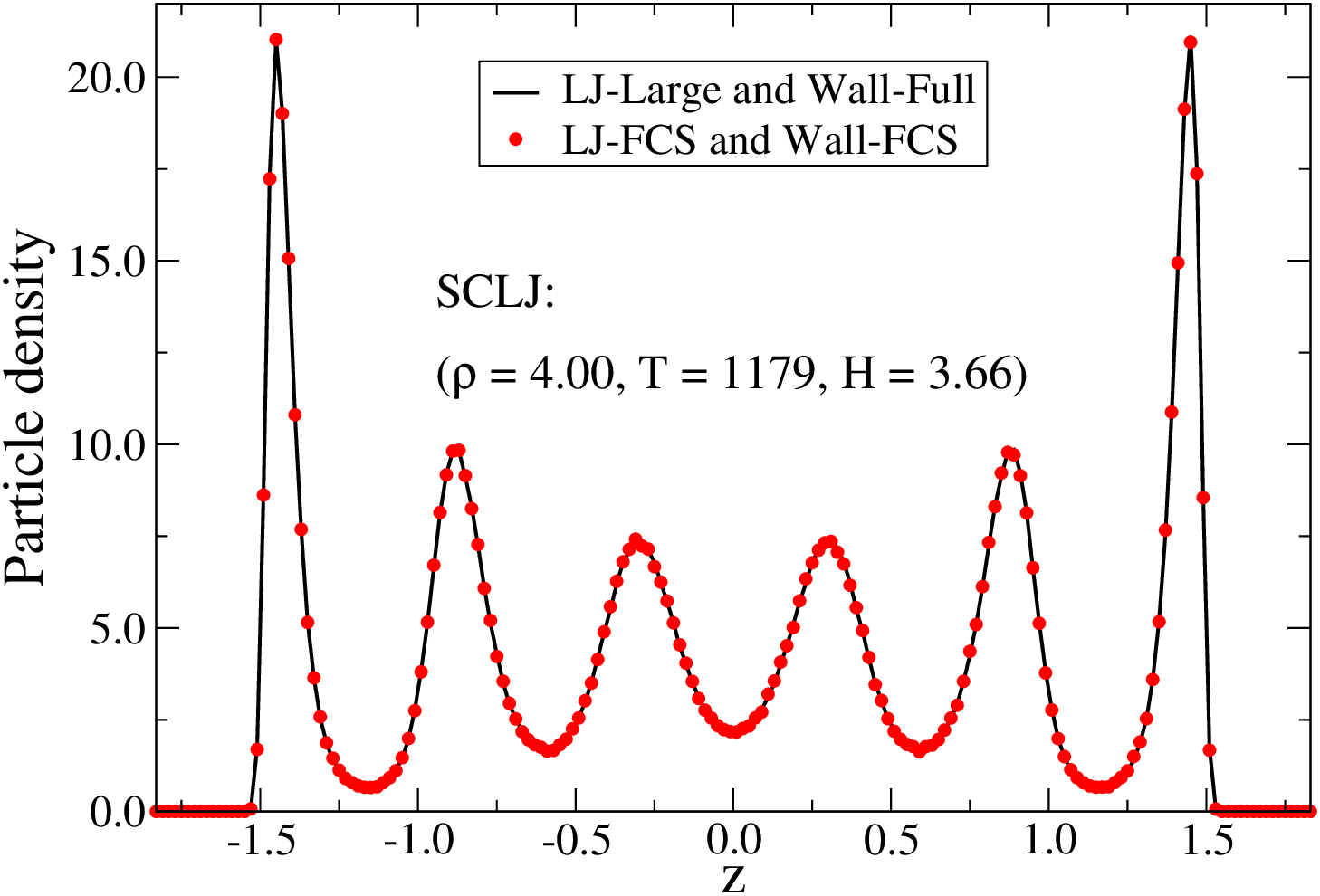}
  \caption{FCS simulations for the SCLJ liquid in a slit-pore at $\rho$ = 4.00, $T$ = 1179, $H$ = 3.66. At 
    this state point the FCSs are pushed to very small distances ($r_{ll,FCS}$ = 0.934 and $r_{wl,FCS}$ = 0.669) well below the 
    potential minima ($2^{1/6}\sigma \approx 1.12$ and $(2/5)^{1/6}\sigma_{w} \approx 0.86$).}
  \label{FCSCON6}
\end{figure}

Figure \ref{FCSCON6} presents results for the confined SCLJ 
liquid at a very high density state point. The distances of the potential 
minima are, respectively, for the liquid-liquid and wall-liquid interactions $2^{1/6}\sigma \approx 1.12$ and $(2/5)^{1/6}\sigma_{w} \approx 0.86$.
The FCS cutoffs are, respectively, $r_{ll,FCS}$ = 0.934 and $r_{wl,FCS}$ = 0.669 and 
thus much smaller than the potential minima. A perfect agreement with the reference simulation is nevertheless obtained confirming that 
the FCS is indeed the fundamental distance also for the wall-liquid interactions.\newline \newline

\section{Conclusion}

The results presented here suggest that bulk and nanoconfined liquids are more similar than what is traditionally believed to be the case. The physics 
as exemplified from the range of interaction is simply as in the bulk liquid; as long as the liquid remains Roskilde simple in confinement. 
The current study has thus detailed a microscopic similarity between some bulk and nanoconfined liquids in terms of the role of the FCS interactions. It is, 
however, likely that this observation may also 
explain the macroscopic similarities between bulk and nanoconfined liquids as they seem to be closely linked to Roskilde 
liquids\cite{hsconfined,dumbbellconfinedroughwalls,goel2009,ingebrigtsen2013,ingebrigtsen2013b}.

Recently, Watanabe \textit{et al.}\cite{watanabe2011} showed that the dynamics of a confined fluid system as a 
function of the distance to the walls can be described to a good approximation using the 
magnitude of the medium-range crystalline order\cite{shintani2006} (MRCO). In this perspective, the relevant range of interaction for the wall-liquid interactions is given by 
the correlation length of MRCO, itself, just as in the bulk liquid. The MRCO perspective is thus fully consistent with the FCS picture proposed here.

We applied in this study an external potential derived from a semi-infinite solid continuum of LJ particles. The SCLJ solid is a 
Roskilde system\cite{paper1,paper2}. Additional studies of more diverse external potentials are thus needed to fully 
clarify the FCS conjecture proposed here. It is, however, likely that a requirement of analyticity
on the external potential must be placed - similar to the bulk liquid\cite{prx}. This is, however, not a crucial restriction
since nature is expected to be analytic. In addition, we applied an approximative linear-cancellation
term near the walls which worked very well for many systems. We have, however, also encountered state 
points for the KABLJ mixture where the cancellation was not perfect (not shown). More work in
this direction is needed.

The results presented here may provide a foundation for understanding  more general interfacial phenomena, which are 
sensitive to the wall-liquid interactions such as 
establisment of contact angles on surfaces\cite{ingebrigtsen2007}, interfacial flow\cite{ma2013}, etc. We welcome 
additional simulation studies of a wide spectrum of Roskilde liquids/external potentials to further test the FCS picture proposed here.

\acknowledgments

The center for viscous liquid dynamics ''Glass and Time'' is sponsored by the Danish National Research Foundation via Grant No. DNRF61. We are grateful to John D. Weeks and Richard Remsing for stimulating 
correspondence concerning the role of attractive forces in nonuniform liquids.


\providecommand*{\mcitethebibliography}{\thebibliography}
\csname @ifundefined\endcsname{endmcitethebibliography}
{\let\endmcitethebibliography\endthebibliography}{}

\end{document}